\documentclass[prc,twocolumn,showpacs,superscriptaddress,nofootinbib]{revtex4}
\usepackage{graphicx}
\usepackage{url}
\begin{document}

\title{Large-scale prediction of the parity distribution in the nuclear
level density and application to astrophysical reaction rates}

\author{D. Mocelj}
\altaffiliation[Current address: ]{ifb International AG, Pf\"affikon,
Switzerland}
\author{T. Rauscher}
\thanks{Corresponding author}
\affiliation{
Departement f\"ur Physik und Astronomie, Universit\"at Basel, Basel, 
Switzerland}

\author{G. Mart\'inez-Pinedo}
\affiliation{GSI Darmstadt, Darmstadt, Germany}
\author{K. Langanke}
\affiliation{GSI Darmstadt, Darmstadt, Germany}
\affiliation{Institut f\"ur Kernphysik, Technische Universit\"at
Darmstadt, Darmstadt, Germany}

\author{L. Pacearescu}
\altaffiliation[Current address: ]{Deutsche Thomson-Brandt GmbH,
Villingen Research Lab, Villingen, Germany}
\author{A. Faessler}
\affiliation{Institut f\"ur Theoretische Physik, Universit\"at T\"ubingen,
T\"ubingen, Germany}

\author{F.-K. Thielemann}
\affiliation{
Departement f\"ur Physik und Astronomie, Universit\"at Basel, Basel, 
Switzerland}

\author{Y. Alhassid}
\affiliation{Center for Theoretical Physics, Sloane Physics Laboratory,
Yale University, New Haven, CT, USA}

\begin{abstract}
A generalized method to calculate the excitation-energy dependent parity
ratio in the nuclear level density is presented, using the assumption
of Poisson distributed independent quasi particles combined with BCS occupation
numbers.
It is found that it is crucial to employ a sufficiently large model space
to allow excitations both from low-lying shells and to higher shells
beyond a single major shell.
Parity ratios are
only found to equilibrate above at least 5--10 MeV of excitation energy.
Furthermore, an overshooting effect close to major shells is found where
the parity opposite to the ground state parity may dominate across a
range of several MeV before the parity ratio finally equilibrates.
The method is suited for
large-scale calculations as needed, for example, in astrophysical
applications.
Parity distributions were computed for all nuclei from the proton dripline
to the neutron dripline and from Ne up to Bi. These results were then
used to recalculate astrophysical reaction rates in a Hauser-Feshbach
statistical model. Although certain transitions can be considerably
enhanced or suppressed, the impact on astrophysically relevant reactions
remains limited, mainly due to the thermal population of target
states in stellar reaction rates.
\end{abstract}

\pacs{26.50.+x -- 21.10.Ma -- 24.60.Dr -- 26.30.+k}

\maketitle

\section{Introduction}
\label{sec:intro}
Knowledge of the nuclear level density in general and, more specifically,
at low excitation energies is of
interest for a number of reasons. Comparison to experimental level
densities helps testing nuclear structure models \cite{abf03}.
In reaction theory, the nuclear level density is an
important ingredient both for the determination of the relevant reaction
mechanism and for the calculation of reaction cross sections \cite{rtk97}.
Because of the low effective interaction energies in astrophysical
applications, the level density at low excitation energy
is usually assumed to be
crucial to determine astrophysical reaction rates.

The level density $\rho$ as a function of spin $J$,
parity $\pi$, and excitation energy $E$ can be written as
\begin{equation}
\rho \left( E,J,\pi \right) = \mathcal{P}\left( E,\pi \right)
\mathcal{F}\left( E,J\right) \rho_\mathrm{tot} \left( E \right) \quad ,
\end{equation}
with the spin projection $\mathcal{F}$.
In most previous applications to astrophysics (e.g.\ \cite{rtk97}),
equally distributed
parities with $\mathcal{P}(E,\pi=\pm 1)=1/2$ for both even and odd
parity have been assumed at all energies but it is obvious that
this assumption is not valid at low excitation energies.
This work focuses on the determination of the parity
projection factor $\mathcal{P}$ and its implication on the calculation
of reaction rates for astrophysics.

In astrophysical applications usually different aspects are emphasized
than in pure nuclear physics investigations. Explosive burning produces
a large number of unstable nuclei for which experimental information is
unavailable. Thus, the study of such models requires prediction of
nuclear properties for a large number of nuclei, several thousands when
considering all nucleosynthesis processes. On the other hand,
reaction rates are obtained by energy-averaging cross sections in the
relevant energy range (see Eq.\ \ref{eq:rr}),
given by the so-called Gamow window \cite{rolfs}.
Therefore, small deviations from the true cross sections may cancel out.
This also may dampen the impact of variations in the predicted
nuclear properties, such as the nuclear level density.

In principle, the nuclear level density should
be extracted from microscopic models.
However, conventional shell model calculations of level density are limited
to the mass range $A\alt 50$ \cite{b01,hkz03,hkz04}
because  of the combinatorial increase of the dimension of the model space
with the number of single-particle levels and/or the number of valence
nucleons. Large-scale shell model calculations of level densities are possible
in the framework of the shell model Monte Carlo (SMMC) method
\cite{dkl95,naal97,aln99}. Most SMMC calculations have been carried out in one full
major shell. However, they can be extended to higher excitation energies by
including all other shells within a mean-field approximation \cite{abf03}.
The SMMC calculations are in general time-consuming and are difficult to
carry out for the large number of nuclei required in large-scale astrophysical
applications so far. Although computers have
become faster, a consistent microscopic description of all required
properties for all nuclei is still not feasible.
This is, of course,
also due to the still insufficient knowledge of nuclear interactions
in general and its effects in neutron- or proton-rich nuclei.
Therefore, many approaches for calculating astrophysical
reaction rates make use of more phenomenological treatments. This does
not imply that they are just mathematical fits to a given property, made
in the region of stability and then extrapolated to unstable nuclei.
Instead, they are based on some physical insight in the origin of the
given property which makes it possible to extend its application also to
unknown nuclei with only few parameters to be adapted. Among the traditional
examples of such approaches is the shifted Fermi-gas model \cite{bethe,cameron}
of the nuclear level density and its variations.
As long as nucleons can be assumed to exhibit a Fermi-gas behavior,
the energy dependence of the level density can be described by a few
parameters. This has been proven repeatedly by comparison to
experimental data but also in comparison to microscopic calculations
\cite{langanke06,abf03,zuk01,isa02}.
Also the behavior of its parameters can be determined by
applying known physical facts and/or by extracting them from different
types of calculations \cite{rtk97,rauapjs}.
Another example of the combined phenomenological
approach is the macroscopic-microscopic Finite Range Droplet Model
(FRDM) \cite{frdm} which still proves to be the most successful model
to predict nuclear masses.

In the spirit of the above we present a macroscopic-microscopic method to
determine the parity ratio of nuclear levels from low to high excitation
energy which can be easily applied to a large number of nuclei.
Providing the
factor $\mathcal{P}$ as a function of excitation energy -- as we do in
the following -- has the additional advantage that it can be combined
with any total level density $\rho_\mathrm{tot}$ from any other approach
to determine the density of odd and even parity states. The method is
based on a quasi-particle model proposed in Ref.\ \cite{abln00} which we extend
by allowing odd particle numbers, inclusion of all shells up to the
11 $\hbar \omega$ oscillator shell, and including excitations between all
considered shells.
Preliminary results were reported in \cite{mocws,mocnpa1,mocnpa2,mocjpg}.

The method is introduced in Sec.\ \ref{sec:ratios}, with a
discussion of some interesting effects found in the parity ratios
included in Sec.\ \ref{sec:ratio_results}. Sec.\ \ref{sec:astro} then
focusses on the astrophysical relevance of the parity distribution.
Astrophysical rates
computed with the new parity ratio are compared to standard rates widely
used in astrophysical applications in Sec.\ \ref{sec:astro_results}. The
paper is concluded with a summary in Sec.\ \ref{sec:sum}.

\section{Determination of parity ratios}
\label{sec:ratios}

\subsection{Basic approach}
\label{sec:ratio_basics}
We start from the assumption of statistically independent particles at
finite temperature.
Because single particle (s.p.) levels can either be occupied or empty, the
probability distribution for the occupation can be assumed to be binomial
\cite{zuk01}.
The probability to occupy $n$ out of $r$ levels is therefore given
by:                                                                             \begin{equation}
B(r;n) = {r \choose n} p^n (1-p)^{r-n},
\end{equation}
where the probability of occupying a level is denoted by $p$. 
We follow \cite{abln00} by replacing the binomial
distribution by a Poisson distribution $P(s)$.
This approximation holds provided the number of levels $r$ is large and
the probability $p$ is small, with the product $rp$  finite.
This can always be achieved by clustering the single particles into
two groups according to their parity and
counting only particles in the group having the opposite parity to the
last occupied level.
The probability to find $n$ particles in that group is then
\begin{equation}\label{ch3:eqn.6a}
P(n) = \frac{f^n}{n!}e^{-f},
\end{equation}
where the average number of particles in the group is given by $f$.
Obviously, neutrons and protons have to be treated separately.
For an even number of nucleons the probability to find the whole system
in a positive parity state is therefore given by
\begin{equation}\label{ch3:eqn.7}
P^+ = \sum_{n,\mathrm{even}} \frac{f^n}{n!}e^{-f} = \cosh f e^{-f},
\end{equation}
and to find the system in a negative parity state by
\begin{equation}\label{ch3:eqn.8}
P^- = \sum_{n,\mathrm{odd}} \frac{f^n}{n!}e^{-f} = \sinh f e^{-f} .
\end{equation}
The probabilities can be related to the total partition function 
\begin{equation}
\label{eq:totpart}
Z=Z^++Z^-
\end{equation}
and the partition functions for odd and even parity states,
$Z^-$ and $Z^+$, by
\begin{eqnarray}
  P^+ &=&  \frac{Z^+}{Z} ,  \nonumber \\
  P^- &=&  \frac{Z^-}{Z} . 
\end{eqnarray}
This leads to the expression used in \cite{abln00} for even-even nuclei:
\begin{equation}
\frac{Z^-}{Z^+} = \tanh f .
\end{equation}
It can be shown \cite{mocthesis} that it can be completely
generalized for even and odd numbers of particles by denoting
the total ground state parity of a nucleus by $g$ and
the opposite parity by $s$:
\begin{equation}
\label{eq:zratio}
\frac{P^s}{P^g} = \frac{Z^s}{Z^g} = \tanh f' ,
\end{equation}
where $f'=f_\mathrm{n}+f_\mathrm{p}$ is now computed from the
sum of the individual average 
particle numbers $f$ of neutrons and protons.

In order to determine the desired parity factor $\mathcal{P}$ the ratio
of the level densities $\rho_g$, $\rho_s$
with different parity has to be known. Applying the
well-known Laplace transform of the partition function and
employing the saddle point approximation, the ratio for an excitation
energy $E$ is given by \cite{mocthesis}
\begin{equation}
\label{eq:rhoratio}
\frac{\rho_s}{\rho_g}= \frac{\beta^s}{\beta^g}\frac{Z^s}{Z^g}
\sqrt{\frac{C^g}{C^s}}e^{(\beta^s-\beta^g)E} ~,
\end{equation}
with the heat capacities $C^g$, $C^s$. In general,
because of the different number of
particles in each group, the inverse nuclear temperatures
$\beta^s$, $\beta^g$ will be different for the same excitation energy $E$.
The required partition functions are determined by
\begin{eqnarray}
Z^g &=&  \frac{1}{1+\tanh f'} ,\nonumber \\
Z^s &=& \frac{1}{1+\frac{1}{\tanh f'}} ~.
\end{eqnarray}
The thermal energy $E$ and the heat capacities can be derived from the
standard thermodynamic relations
\begin{eqnarray}
\label{eq:thermodyn}
E &=& - \frac{\partial\ln Z}{\partial \beta} \nonumber \\
C &=& -\beta^2\frac{\partial^2\ln Z}{\partial \beta^2} ~.
\end{eqnarray}

Finally, the parity projection factor $\mathcal{P}$ is given by
\begin{eqnarray}
\mathcal{P}(E,\pi=\pi_g)=\mathcal{P}_g &=& \frac{\rho_g}{\rho_\mathrm{tot}} =
\frac{1}{1+\xi} \, , \nonumber \\
\mathcal{P}(E,\pi=\pi_s)=\mathcal{P}_s &=& \frac{\rho_s}{\rho_\mathrm{tot}} =
\frac{1}{1+\frac{1}{\xi}} \, ,
\end{eqnarray}
with
\begin{equation}
\xi=\frac{\rho_s}{\rho_g}
\end{equation}
and $\pi_g$ being the ground state parity of the nucleus while $\pi_s$
being the opposite parity.

\subsection{Determination of the mean occupation number}
\label{sec:f}
The central quantities $f_\mathrm{n}$, $f_\mathrm{p}$ were
determined separately for neutrons and protons. In the following, we denote
them by $f$ for simplicity and imply that the two kinds of nucleons were
treated separately with their respective s.p.\ levels and particle numbers.
Assuming Fermi-Dirac distributed particles,
the occupancy $f_k$ of each s.p.\
level with energy $\epsilon_k$ given by
\begin{equation}
f_k^{FD} = \frac{1}{1+e^{\beta(\epsilon_k-{\bar \mu})}}
\end{equation}
and the average number of particles by
\begin{equation}
\langle n \rangle = \sum_k f_k ~.
\end{equation}
Proceeding as before by dividing the s.p.\ levels into a group
exhibiting the same parity $\pi_g$ as the last occupied state below the
chemical potential $\bar \mu$ and another group with opposite parity $\pi_s$,
the mean value $f$ in the Poisson distribution is then given by
\begin{equation}
\label{eq:fFD}
f = \sum_{k\in\pi_s} f^{FD}_k = \sum_{k\in\pi_s} \frac{1}{1+e^{\beta(\epsilon_k
-{\bar \mu})}} ~.
\end{equation}
The chemical potential is found iteratively from the particle number
equation
\begin{equation}
  n = \sum_k \frac{1}{1+e^{\beta(\epsilon_k-{\bar \mu})}} \quad .
\end{equation}

With decreasing nuclear temperature $T_\mathrm{nuc}$ pairing interactions
become increasingly important in nuclear systems, leading to nucleon
pairs in the ground state and at low excitation energies.
The breaking of such pairs requires additional energy.
Hence, the Fermi-Dirac distribution will not be
able to describe the occupation properly at high values of
$\beta=1/T_\mathrm{nuc}$.
For example, this can be seen in Fig.\
3 of \cite{abln00} where a comparison between Fermi-Dirac and BCS
occupancies and resulting partition ratios $Z_-/Z_+$
are shown for even-even nuclei.
At high nuclear temperatures
the microscopic distributions are well described by Fermi-Dirac statistics
but for lower temperatures deviations start to appear. For those cases of
even-even nuclei, the probability to find an even number of
particles in the $\pi_s$ parity group is clearly enhanced,
odd numbers of particles are suppressed as seen in the
comparison to the SMMC result in \cite{abln00}.
Similar effects were reported in \cite{abf03}.

Pairing effects can be included by using the well-known
Bardeen-Cooper-Schrieffer (BCS) formalism \cite{bcs}, introducing 
the BCS occupancy for quasi-particles
\begin{equation}
f_k^{BCS} = 1 / (1+\exp(\beta E_k))
\end{equation}
with the
quasi-particle energy 
\begin{equation}
E_k=\sqrt{(\epsilon_k-{\bar \mu})^2+\Delta^2}~.
\end{equation}
The chemical potential ${\bar \mu}={\bar \mu}(\beta)$ and the pairing gap $\Delta=\Delta(\beta)$ are
determined for each nuclear temperature $T_\mathrm{nuc}=1/\beta$
by solving a non-linear
system of equations, the well-known particle number and gap equation
\begin{eqnarray}\label{eq:bcs}
  n &=& \frac{1}{2}\sum_k\left(1-\frac{\epsilon_k-{\bar \mu}}{E_k}
    \tanh\left(\frac{\beta}{2}E_k\right)\right) \nonumber \\
  \frac{2}{\Delta} &=& \frac{G}{2}\sum_k \frac{\Delta}{E_k}\tanh\left(\frac{\beta}{2}E_k\right) ~,
\end{eqnarray}
where $G$ is the usual effective pairing coupling constant which
determines the zero-temperature pairing gap $\Delta$.
Although the single particles are no
longer independent in the paired solution, 
the Poisson
approach of Sec.\ \ref{sec:ratio_basics} can be kept, assuming that the
quasi particles are statistically independent here.
For the calculation of the mean occupation in the $\pi^s$ parity group,
it is sufficient to consider only
contributions from quasi-particles, as condensed pairs do not contribute
to any parity change of the system.
This leads to
\begin{equation}
\label{eq:fBCS}
f = \sum_{k\in\pi_s} f^{BCS}_k = \sum_{k\in\pi_s} \frac{1}{1+ \exp(\beta E_k)} ~.
\end{equation}
It should be noted that here the chemical potential $\bar \mu$ and the pairing
gap $\Delta$ are temperature-dependent, i.e.\ have to be known as a function
of $\beta$ or $T_\mathrm{nuc}$. The BCS equations are solved iteratively
for each inverse temperature $\beta$.

At sufficiently high nuclear temperature, pairing becomes negligible and
there is a phase transition from the BCS to the Fermi-Dirac regime.
However, this transition is not smooth when only using the above
description because
the BCS equations break down above a critical temperature $T_\mathrm{c}=1/\beta_\mathrm{c}$.
Due to the phase transition occuring at $T_\mathrm{c}$, the derivatives of
the partition function $Z$ (containing the mean occupancy $f$)
are discontinuous when moving from the BCS
regime to the Fermi-Dirac regime. This induces numerical problems in the
solution of Eq.\ (\ref{eq:rhoratio}) via Eq.\ (\ref{eq:thermodyn}),
especially also because the
different parity groups exhibit different temperatures. To avoid the
problematic temperature range would be too limiting because the range
of excitation energies most interesting for studying the parity ratios
would be excluded. Therefore, we employed an approximation by extrapolating
the pairing gap to higher nuclear temperatures. In fact, the transition from
the BCS to the Fermi-Dirac regime should be smooth in nuclei because the pair
coherence length is much larger than the nuclear radius \cite{la01}. In such
systems with dimensions smaller than the coherence length, fluctuations in
the order parameters become important and wash out the discontinuity.
\begin{figure}
\includegraphics*[width=\linewidth,viewport= 0 0 596 422]{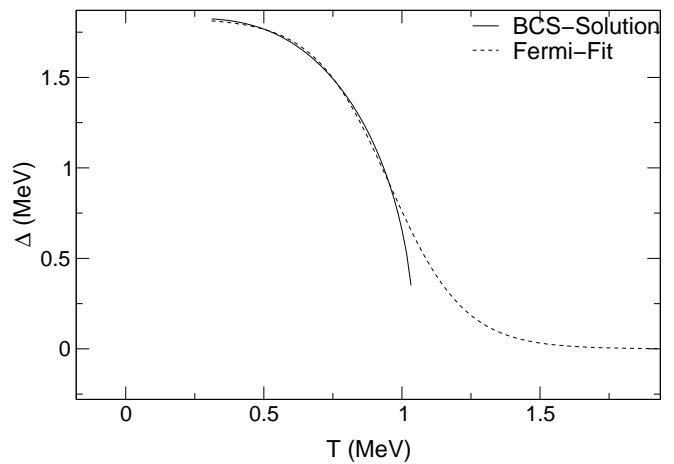}
\caption{\label{fig:fermifit}
Dependence on nuclear temperature of the
proton pairing gap $\Delta$ in $^{66}$Zn and its fit
by a Fermi-type function (see text).}
\end{figure}

As the pairing effects vanish for high nuclear temperatures, the pairing gap
itself should approach a value of Zero, thus recovering the Fermi-Dirac
shape of the distribution (see, e.g., Ref.\
\cite{kh05} for experimental indications
of this effect). This behavior of the gap could
be rigorously obtained, for instance, by using the
static path approximation, integrating over the static fluctuations
of the gap \cite{chal07}. However, as has been shown in SMMC calculations \cite{dkl95,ldrk96},
this behavior can be simply parameterized in terms of
a Fermi-type dependence:
\begin{equation}
\label{eq:fermifit}
\Delta^\mathrm{fit}(\beta)=\Delta^\mathrm{fit}(T_\mathrm{nuc})=\Delta_T=\frac{\Delta_0}{1+\exp(-(T_\mathrm{nuc}-T_\mathrm{nuc}')/a)} \ ,
\end{equation}
where $\Delta_0$ denotes the pairing gap for $T_\mathrm{nuc}=0$.
The parameters $T_\mathrm{nuc}'$ and $a$ were fitted
so that the function approximated as closely as possible
the behavior of the BCS gap below the critical temperature.
As starting values we chose $T_\mathrm{nuc}'=0.8 T_\mathrm{c}$
and $a=(T_\mathrm{nuc}'-T_\mathrm{nuc})/\ln(\Delta_0/\Delta^\mathrm{BCS}_{T_\mathrm{nuc}'}-1)$
which yielded a good
fit to the BCS pairing gap $\Delta^\mathrm{BCS}=\Delta$
below $T_\mathrm{nuc}'$.
This results in a suppression of the pairing gap across a
temperature range of the order of MeV, as also previously found in
\cite{dkl95,ldrk96,langanke06}.
For illustration,
Fig.\ \ref{fig:fermifit} shows a comparison of the BCS solution and the
fit for $^{66}$Zn.
The results are not sensitive to small variations in the choice of the
starting values.
These fits were subsequently used in the calculation
of the chemical potential $\bar \mu$ and
finally of the mean occupation number $f$
by applying Eq.\ (\ref{eq:fBCS}) for all values of $\beta$.

\subsection{Input quantities and consistency}
\label{sec:ratio_inputs}
The parity projected partition functions $Z^g$, $Z^s$
are related to each other, to the
total partition function $Z$, and to the average occupancy $f$
by Eqs.\ (\ref{eq:totpart}) and (\ref{eq:zratio}). Thus, the total partition
function $Z$ and the average occupation number $f$ remain to be
determined.

The determination of $f$ as described in the previous section implicitly
requires knowledge of the pairing strength $G$ and the s.p. levels.
For systems with an even number of particles,
the starting value $\Delta_0$ of the pairing gap
was extracted from odd-even mass differences as described in \cite{rtk97}.
This value was then used to determine the effective pairing strength $G$
by solving the coupled Eqs.\ (\ref{eq:bcs}) for $T_\mathrm{nuc}=0$.
Further application of Eqs.\ (\ref{eq:bcs}) with fixed $G$
at finite nuclear temperatures
yields the temperature-dependence of $\Delta$ and the
required occupation of quasi-particle states.
Differently from the treatment
of the pairing gap $\Delta_0$ in \cite{rtk97}, here we set $\Delta_0=0$
for systems with odd numbers of particles because the pairing effects are
expected to be weak and quickly vanishing with increasing $T_{nuc}$.
This means that a pure Fermi-Dirac distribution is used for odd particle numbers
without the need to fit the phase transition.

The s.p.\ levels were calculated in a
deformed Saxon-Woods potential \cite{dpps69} with parameters from \cite{tops79}
which reproduce experimental data well \cite{nfs93,smnf93}. The same
deformation was used as in \cite{rtk97,rt00,rt01}, taken from \cite{frdm}.

\begin{figure}
\includegraphics[width=\linewidth]{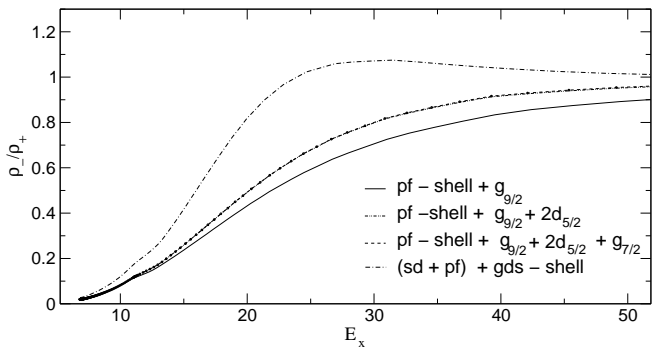}
\caption{\label{fig:shells}Ratio of the level densities with odd and
even parities as a function of excitation energy $E_\mathrm{x}$ in MeV
in the nucleus $^{56}$Fe, calculated with different sizes
of the model space. It is evident that it is not sufficient to only
include the $pf+g_{9/2}$-shell.}
\end{figure}

Fig.\ \ref{fig:shells} illustrates the dependence of the parity distribution
on the single-particle model space in $^{56}$Fe.
For this nucleus, the addition of the $2d_{5/2}$ orbital to the
$pf+g_{9/2}$-shell leads to a deviation over 10\% in the parity ratio
above excitation energies $E_x$ of about 15 MeV.
Further inclusion of higher states such as $1g_{7/2}$ does not
change the result for $^{56}$Fe (the lines for the results overlap in the
figure) but will be important for heavier nuclei.
Therefore we had to include all major shells up to 11 $\hbar \omega$.

We also see in Fig.\ \ref{fig:shells} that levels usually considered as
inert core can also have an important effect.
This is clearly seen in the result
accounting for excitations across the full $sd+pf+gds$-shells. Here we find
deviations of at least 10\%
already at $E_x \simeq 7$ MeV and larger deviations at higher excitation
energies. This is mainly due to excitations from the $sd$-shell.

As a further ingredient, a total nuclear level density $\rho_\mathrm{tot}$
is required
in the determination of the total partition function $Z=Z(\beta,\rho_\mathrm{tot})$. In
principle, any total level density determined in any approach could be
supplied here.
Since we want to apply the derived parity factor to the level density
of \cite{rtk97}, we took the total level density and the level density
parameters from there.

\begin{figure}
\includegraphics[width=6cm,angle=-90]{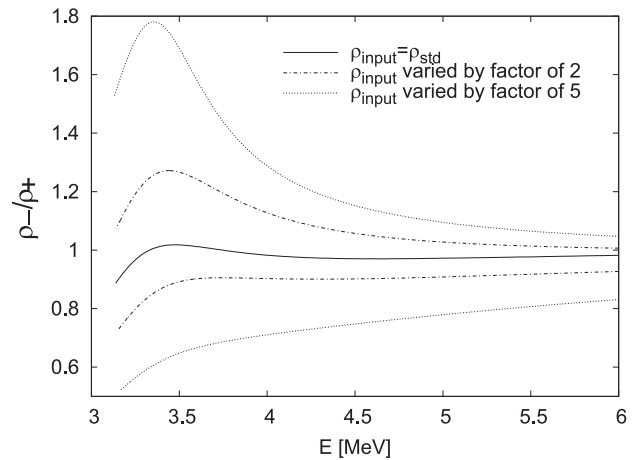}
\caption{\label{fig:consistency}
Influence of a variation of the input level density $\rho_\mathrm{input}$
on the calculated
parity ratios in $^{70}$Zn. The full line is
obtained by using the standard level density $\rho_\mathrm{std}$
\cite{rtk97}.
The dashed and dotted lines correspond to variations of the
input level density by factors 2 and 5, respectively. The s.p.\ levels are
held fixed.}
\end{figure}

In a fully consistent model, s.p.\ levels determine the level density
assuming all effects have been properly
accounted for and provided that the level density can be computed, e.g.\ in a
shell model, for all nuclei at all relevant
excitation energies. In principle, this can be achieved, e.g., within the
approach of Ref.\ \cite{abf03}, but is difficult to carry out for a large
number of nuclei.
For practical purposes it is therefore necessary to include
level densities derived in other approaches, such as
the shifted Fermi-gas. Proceeding in this manner, however, great care has to be
exercised to achieve consistency between
the s.p.\ structure, in our case obtained from the Saxon-Woods model,
and the input level density.
Inconsistency will lead to spurious
effects in the parity ratio as shown in Fig.\ \ref{fig:consistency}.

Realistically one has to expect that both the s.p.\ level structure from
the Saxon-Woods potential and our total level density bear inherent
uncertainties. Thus, when striving for consistency it is anticipated that both
inputs should be adapted. Here, we used an iteration method which only varies
the level density and not the Saxon-Woods potential, assuming that the s.p.
structure is correct. Although this does not limit the applicability of
our results, it has to be emphasized that the variation of the level
density $\rho_\mathrm{input}=\rho_\mathrm{tot}$
shown in Fig.\ \ref{fig:consistency}
and found in our iteration procedure is not due
to the uncertainty of $\rho_\mathrm{tot}$ alone but rather is supposed
to contain the combined uncertainties in $\rho_\mathrm{tot}$ and
the s.p.\ levels.

In each iteration step the parity projected level-densities
\begin{eqnarray}
  \rho_g(E) &=& \frac{1}{\sqrt{2\pi C^g}}\beta^g \exp(\beta^g E + \ln Z^g) ~, 
\nonumber \\
  \rho_s(E) &=& \frac{1}{\sqrt{2\pi C^s}}\beta^s \exp(\beta^s E + \ln Z^s)
\end{eqnarray}
were calculated with an input level density $\rho_\mathrm{input}$.
The total level density in this step is then
given by the sum of the parity projections:
\begin{equation}
  \rho^{({{\rm{it}}})} = \rho_g + \rho_s ~.
\end{equation}
In the first iteration step the standard
level density from Ref.\ \cite{rtk97} was used for $\rho_\mathrm{input}$.
In each following step, $\rho^{({\rm{it}})}$ of the previous step
became the new $\rho_\mathrm{input}$ and the procedure continued until
$\rho^{({\rm{it}})}$ and $\rho_\mathrm{input}$ converged.
Convergence was usually achieved quickly, typically in the third iteration step.
\begin{figure}
\includegraphics[width=\linewidth]{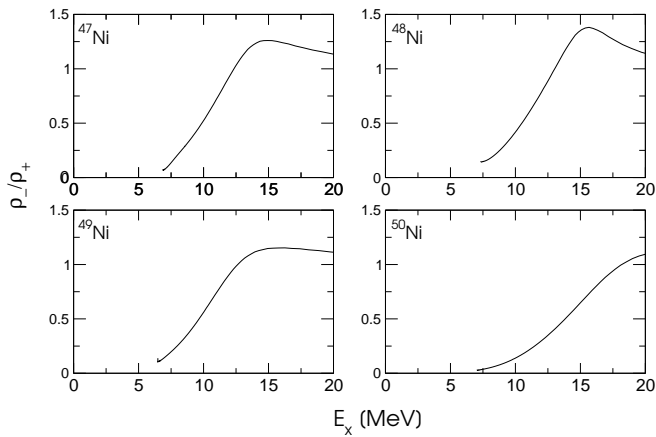}
\caption{\label{fig:ni}Ratio of the parity projected level densities
as a function of excitation energy for Ni isotopes around
the neutron $sd$-shell closure.}
\end{figure}

\subsection{Results and discussion of selected examples}
\label{sec:ratio_results}
Using the approach described in the previous sections,
we have calculated ratios of parity-projected level densities for all nuclei
from Ne to Bi and from proton dripline to neutron dripline,
adopting the large 11$\hbar\omega$ model space. All target nuclei
in \cite{rt00,rt01} are covered\footnote{It should be noted that this also
impacts, for example, the resulting driplines as the same inputs as
in these references were used here.}. The results can be obtained from the
American Institute of Physics' EPAPS \cite{EPAPS} or directly from the authors
at \url{http://nucastro.org}.

\begin{figure}
\includegraphics[width=\linewidth]{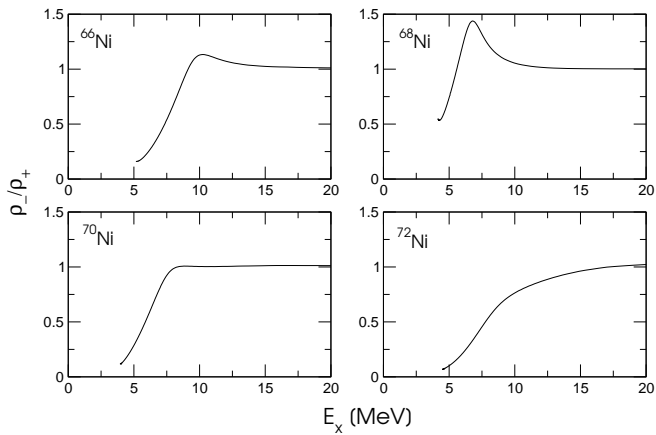}
\caption{\label{fig:ni68}Ratio of the parity projected level densities
as a function of excitation energy for Ni isotopes around the neutron
$pf$-shell closure.}
\end{figure}

\begin{figure}
\includegraphics[width=\linewidth]{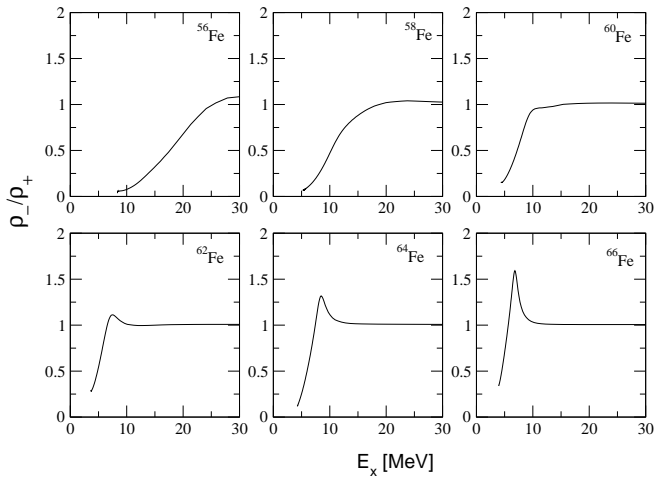}
\caption{\label{fig:fechain}
Ratio of the parity projected level densities as a function of
excitation energy for Fe isotopes up to $N=40$.
}
\end{figure}

\begin{figure}
\includegraphics[width=\linewidth]{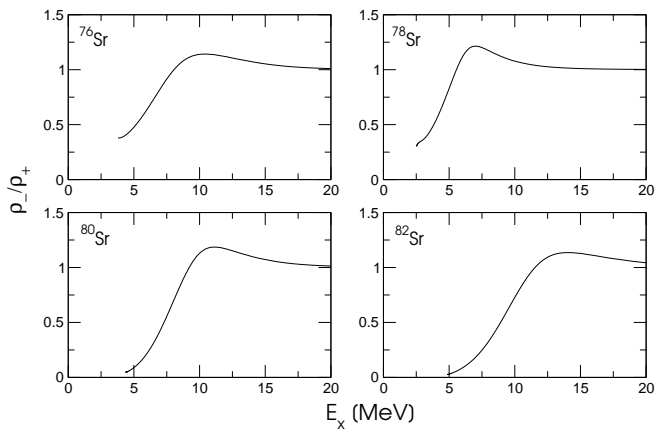}
\caption{\label{fig:sr78}
Evolution of the ratios of the parity projected level densities as a
function of excitation energy for Sr isotopes around the neutron $pf$-shell
closure.
}
\end{figure}

\begin{figure}
\includegraphics[width=\linewidth]{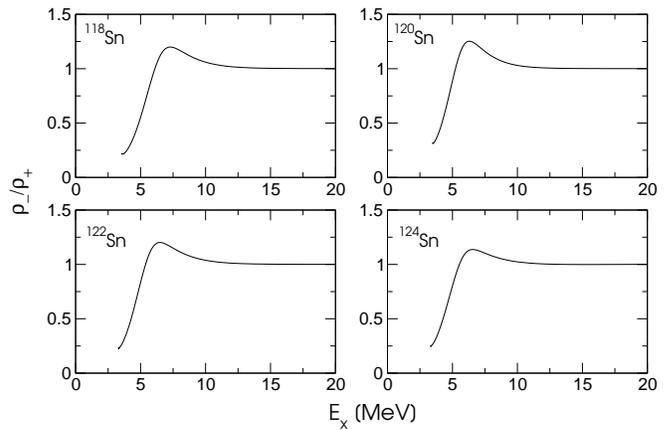}
\caption{\label{fig:sn120}
Evolution of the ratios of the parity projected level densities as a
function of excitation energy for Sn isotopes around the neutron
$gds$-shell closure.
}
\end{figure}

\begin{figure}
\includegraphics[width=\linewidth]{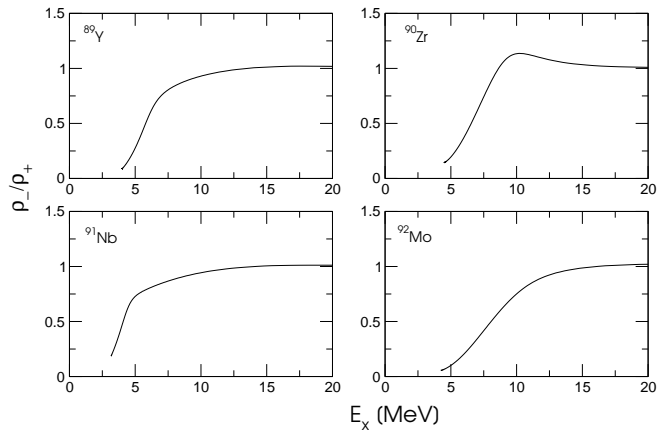}
\caption{\label{fig:n50iso}
Evolution of the odd- to even-parity ratio within the $N=50$ isotones in the
vicinity of the proton $pf$-shell closure.
}
\end{figure}

Representative for our results are Figs.\ \ref{fig:ni}--\ref{fig:n50iso}.
It can immediately be noticed that the parities equilibrate only at
excitation energies above about $5-10$ MeV or even higher (see, e.g.,
Fig.\ \ref{fig:ni}), even for the more heavy nuclei.
This underlines the importance of using a parity projection factor
$\mathcal{P}\neq 1/2$ at most energies of astrophysical interest.

Surprising at first glance is the effect we found at closed shells.
The parity ratio can overshoot the equilibrated value of $\rho_s/\rho_g=1$
by 20--60 \% before finally reaching the equilibrium value at higher
excitation energies. The overshooting can be understood by the fact that
parity can only be changed by excitations between s.p.\ levels of different
parity. When an oscillator shell is completely filled, i.e.\ the next available
subshell exhibits opposite
parity, any excitation will change the parity of the system, resulting in
a dominance of opposite parity states.

While studying the following examples, it should be kept in mind that
both neutron and proton excitations contribute to the parity ratio. When
comparing the ratios within an isotopic chain, however, the proton
contribution will remain the same and differences between isotopes can be
attributed to the change in neutron number. Similar considerations apply
to the comparison of isotones and the changing proton number.
For example, Fig.\ \ref{fig:ni}
shows the evolution of the parity ratio
in Ni isotopes around the neutron $sd$-shell closure.
The $sd$-shell is filled completely with neutrons
for $^{48}$Ni. Every excitation of $^{48}$Ni populates a single particle
level of opposite parity leading
to a maximal parity change. The formation of the peak can already be seen for
$^{47}$Ni but the position of
the peak is shifted towards smaller excitation energies.
This energy shift is about 1 MeV reflecting the
energy that is needed to break a neutron pair prior to excitation.
The isotopes $^{49}$Ni and $^{50}$Ni, which both already
populate the $pf$-shell in neutrons, equilibrate at much higher excitation
energies.  Parity change can be achieved by excitations
to the $g_{9/2}$ shell or by excitations from the $sd$ to the $pf$ shell,
requiring higher energy on average than for the preceding isotopes.

The evolution of the parity ratio for Ni isotopes at the interface
of the $pf$ and $gds$ shell is shown in Fig.\
\ref{fig:ni68}. Again, a similar behavior as described above
can be observed. The $N=40$ neutron shell is
completely filled for $^{68}$Ni. Each excitation from the last occupied
$2p_{1/2}$ level with negative
parity will populate levels from the $gds$ shell with positive parity
resulting in a parity-change of
maximal amplitude. The formation of the peak can already be seen for $^{66}$Ni.
The position of the peak
is shifted to higher energies as a larger gap between the $2f_{5/2}$ level
and the $g_{9/2}$-level has to be bridged.

The transitions between the same shells as above in Fe isotopes is shown in
Fig.\ \ref{fig:fechain}.
Moving from $^{56}$Fe, where the neutron $pf$-shell is only half-filled,
to $^{66}$Fe,
where the $2p_{1/2}$-shell is completely filled, it can be seen that the
ratio approaches unity for lower
values of the excitation energy as one approaches the $N=40$ shell closure.
As the parity can only be changed
by excitations either from the $sd$ to the $pdf$ shell
or from the $pf$ to the $gds$ shell, the ratio will equilibrate
faster with increasing neutron number
as the gap between the last occupied orbit in the $pf$-shell and the
$gds$-shell will decrease. For $^{66}$Fe,
where the $pf$ shell is completely filled, a pronounced peak around 8 MeV
occurs.

A similar behaviour is illustrated by Fig.\ \ref{fig:sr78}
for the evolution of the
parity ratio in Sr isotopes at the $pf$ and $gds$ interface. Again, the
ratio reaches a maximum in $^{78}$Sr with the filling of the neutron
$pf$-shell. A comparison
to the Fe isotopes with the same neutron numbers in Fig.\ \ref{fig:fechain}
shows the impact of deformation. The FRDM model \cite{frdm} predicts
the Sr isotopes shown here to be strongly deformed. The resulting
level splitting leads to a wider, less pronounced peak in the Sr
isotopes as compared to the Fe isotopes.

Another interesting case is seen in Fig.\ \ref{fig:sn120} which displays
the evolution of the parity ratio in Sn isotopes in the vicinity
of the $gds$- and $pfh$-shell.
The $h_{11/2}$-subshell is half filled at $^{120}$Sn
due to the fact that it is located below the $2d_{3/2}$
and $3s_{1/2}$-subshells, the latter two belonging to the $gds$-shell.
Comparatively small excitation energy is required to change the parity
by either moving neutrons from the $h_{11/2}$-subshell to the higher
$ds$-states or by populating the $h_{11/2}$-subshell with neutrons from
lower $gd$-states. The effect of the latter is largest when the
$h_{11/2}$-subshell is least occupied, the effect of the former becomes
larger with increased occupation. That is why there is an overshooting
peak found in all Sn isotopes shown in Fig.\ \ref{fig:sn120}, with
the one in $^{120}$Sn being the most pronounced because both excitations
contribute almost equally. However, due to the mixing of $gds$- and
$h_{11/2}$-states, the overshooting is not as strong as the one found for
$^{66}$Fe at the $pf$-closure.

Above, the evolution of the parity ratio was discussed in isotopic chains,
where each parity change results
from an excitation of neutrons only. A similar effect is found for protons.
For instance, Fig.\ \ref{fig:n50iso} shows the evolution of
 the $N=50$ isotone parity ratio.
The proton numbers for the four shown nuclei (Y, Zr, Nb, and Mo) are
39, 40, 41, and 42, respectively.
Again, a peak can be observed in $^{90}$Zr as the proton
$pf$-shell is completely filled for this nucleus. Weak deformation
is predicted for these nuclei which leads to a suppression of the peak.

\section{Application in astrophysics}
\label{sec:astro}

\subsection{Introduction}

In the following we discuss the implications of a realistic parity
distribution in astrophysical applications.
The astrophysical reaction rate is the central quantity in the nuclear
reaction networks employed to follow nucleosynthesis in different
astrophysical environments. For nucleon-nucleus and nucleus-nucleus
reactions it is given by folding the reaction cross section with
the Maxwell-Boltzmann velocity distribution of the interacting nuclei
\begin{equation}
\label{eq:rr}
r=n_1 n_2
\sqrt{\frac{8}{\pi M \left( kT^{*}\right) ^3}}\int
\limits _{0}^{\infty }\sigma ^{*}(E')E'\exp \left( -\frac{E'}{kT^{*}}\right) dE'\, .
\end{equation}
The stellar plasma temperature is denoted by $T^*$, the reduced mass of
the interacting nuclides by $M$, and their number densities by $n_1$, $n_2$.
The stellar cross section is defined as
\begin{equation}
\label{eq:csstar}
\sigma ^{*}(E_\mathrm{c.m.})={\sum _{\mu }(2J^{\mu }+1)\exp (-E^{\mu }/kT^{*})
\sum _{\nu }\sigma ^{\mu \nu }(E_\mathrm{c.m.})\over \sum _{\mu }(2J^{\mu }+1)
\exp (-E^{ \mu }_{i}/kT^{*})}\quad ,
\end{equation}
including sums over excited states $\mu$, $\nu$ in the target and final
nucleus, respectively, with $J^\mu$ and $E^\mu$ denoting spin and excitation
energy of the $\mu$-th target state.

\subsection{Relevant reaction theory}
\label{sec:astro_models}
For the majority of reactions relevant in astrophysics,
the reaction cross sections $\sigma ^{\mu \nu }$
 have to be calculated with different
reaction models as experimental information often is not available in the
relevant energy range or even impossible to obtain due to the short
half-lives of the involved nuclei.
Even if data is available, it can only provide a measure of
$\sigma ^{\mu}= \sum _{\nu }\sigma ^{\mu \nu }$
and thus does not provide the required $\sigma^*$.
Depending on the number of
resonances in the Gamow window \cite{rtk97},
one has to consider direct reactions,
the influence of few resonances, or an average over many resonances
\cite{dr06}. Here, we want to focus on the latter and on direct capture.

Provided there is a high number of resonances at the relevant interaction
energy, the formation and decay of a compound nucleus can be described
by averaged transmission coefficients in the 
statistical model or
Hauser-Feshbach formalism \cite{HF52}. The transmission coefficients
describing the decay of a compound state $k$ via a process
$y$ -- with $y$ usually being emission of nucleons, $\alpha$
particles, or $\gamma$ quanta --
include sums of the type $^kT_y^{J^k\pi^k}=\sum _{\nu } {^kT_y^{J^k\pi^k
\rightarrow J^\nu \pi^\nu}}$.
Depending on the nucleus, reaction $Q$ value, and interaction
energy $E_\mathrm{c.m.}$, the
properties (spin $J^\nu$, parity $\pi^\nu$, excitation energy $E^\nu$) of
the accessible states
may not be known above a certain energy $E_\mathrm{c}$.
Then the sum is replaced by an integration over a level density
\begin{eqnarray}
^kT_y^{J^k\pi^k}&=&\int \limits _{E_\mathrm{c}} ^{Q+E_\mathrm{c.m.}}
\sum _{J^\nu,\pi^\nu} T_y ^{J^k\pi^k\rightarrow J^\nu \pi^\nu}
(Q+E_\mathrm{c.m.}-E^\nu) \times \nonumber \\
&&\times \rho(E^\nu,J^\nu,\pi^\nu) dE^\nu \, .
\label{eq:trans}
\end{eqnarray}
The integration starts from $E_\mathrm{c}=0$ in most cases far from
stability where no excited states are known. The level density
$\rho(E^\nu,J^\nu,\pi^\nu)$
includes the assumption of the parity distribution. It should be noted
that this is the level density in the final nucleus which is identical
to the compound nucleus only for capture reactions. The interaction
energies $E_\mathrm{c.m.}$
in astrophysics typically are smaller than 1 MeV for neutrons and
5--10 MeV for charged projectiles. A comparison with our results from the
previous section shows that the parity ratio obviously has not
equilibrated at the relevant excitation
energies, even though the reaction $Q$ value
has to be added to these energies.
\begin{figure}
\includegraphics[width=\linewidth]{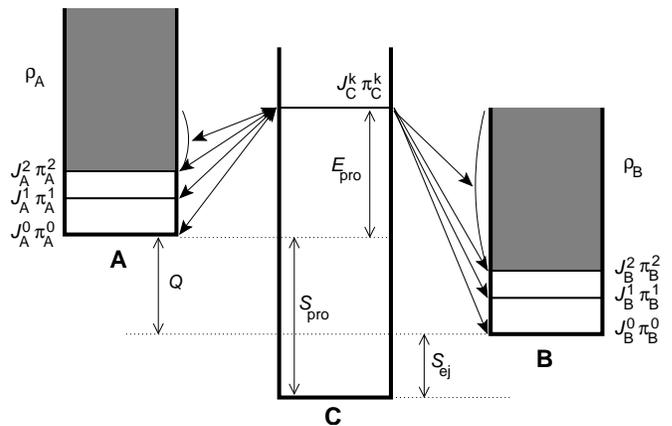}
\caption{\label{fig:CN}
Energy scheme of a compound nucleus reaction $A+\mathrm{pro}\rightarrow
C^* \rightarrow \mathrm{ej}+B$. For a capture reaction on target nucleus $A$
the ejectile "ej" will be $\gamma$ radiation and the final nucleus $B$ will
be identical to the compound nucleus $C$. The standard reaction $Q$ value is the
difference between the separation energies of the projectile and the ejectile
$Q=S_\mathrm{pro}-S_\mathrm{ej}$. When considering individual transitions
between excited states in the target and the final nucleus the released
energy becomes $Q=(S_\mathrm{pro}+E_A^\mu)-(S_\mathrm{ej}+E_B^\nu)$, with
the excitation energies $E_A^\mu$, $E_B^\nu$. The excitation
energy of the accessed compound state is then $E^k_C=S_\mathrm{pro}+E_\mathrm{pro}+E_A^\mu$, with $E_\mathrm{pro}$ being the projectile energy. The decay
of the compound state $k$ is computed by a sum of transitions to individual
states and by integrating over an average level density (shaded area) above
the last included state. The target $A$ can become excited by thermal
population of the excited states in an astrophysical plasma.
}
\end{figure}

Considering thermal excitation of the target states in the transmission
coefficients for the formation of the compound nucleus in state $k$
a relation similar to Eq.\ (\ref{eq:trans})
can be derived for the transmission coefficients $^k T_{y'}$ connecting
target and compound state via a process $y'$,
this time inserting the level density in the
target nucleus. However, due to the exponential suppression
of the population of the excited target states (see Eq.\ \ref{eq:csstar})
significant contributions to the integral will only stem from the 
lowest energies.

The central assumption of the Hauser-Feshbach model is that the compound
state at energy $E^k$ is highly degenerate. Due to the large number of
overlapping resonances, the excited compound nucleus can have any
combination of spin and parity. The possible combinations will be weighted
by the transmission coefficients $^k T_{y'}$ describing the formation
process and, e.g., accounting for spin selection rules. However, in principle
one has to sum over all compound spins $J^k$ and parities $\pi^k$
when calculating the statistical model cross section
\begin{equation}
\label{eq:hfcs}
\sigma_\mathrm{HF}^{y'y} \propto \sum \limits _{J^k\pi^k} \left( 2J^k+1 \right)
\frac{^kT_{y'}^{J^k\pi^k} \;
{^kT_{y}^{J^k\pi^k}}}{\sum \limits _{z=y,y',\dots} {^kT_z^{J^k\pi^k}}} \, .
\end{equation}
The denominator contains a total transmission coefficient
for each state $J^k\pi^k$, including
all possible deexcitation processes, e.g.\
emission of protons, neutrons, $\alpha$ particles, and $\gamma$ quanta.
The scheme of a compound nucleus reaction is sketched in Fig.\ \ref{fig:CN}.

\begin{figure}
\includegraphics[width=\linewidth]{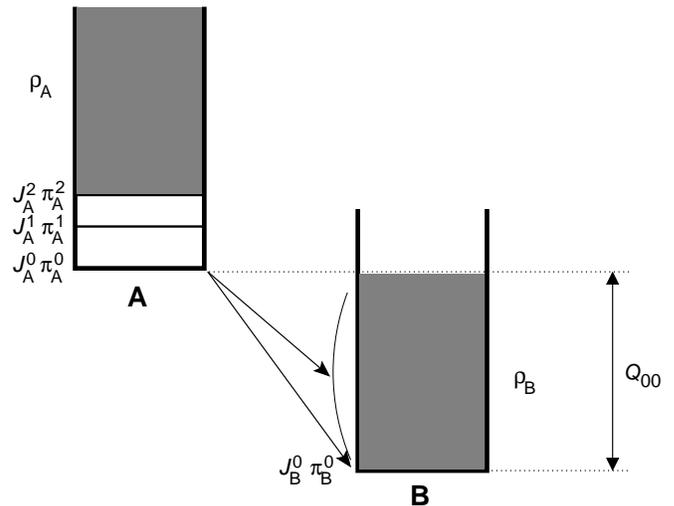}
\caption{\label{fig:DC}
Energy scheme of a direct capture reaction $A+\mathrm{pro}\rightarrow
B+\gamma$. The energy of the emitted $\gamma$ quantum is the energy difference
between initial target state and final state. For instance, $Q_{00}$ is
the released energy for the ground state to ground state transition and is
equal to the binding energy of the projectile in the final nucleus $B$.
As in the case of the compound mechanism,
it is possible to include an integration over a level density (shaded region)
when the exact properties of the nuclear states are not known. For
simplicity, only transitions originating from the ground state of target $A$
are drawn here. For stellar reaction rates it would be important to also
include transitions from thermally excited target states.
}
\end{figure}

Direct reactions proceed in a different manner, without formation of an
excited compound nucleus. A nucleon or nucleon group is directly transferred
from the projectile to the final state without excitation of any other
nucleons in the system \cite{dr06,sat83,glen83}. This mechanism will
dominate at very high interaction energies where the compound nucleus
formation is suppressed due to the short interaction timescale and at low
energies in the absence of resonances, again suppressing the formation of
a compound nucleus. In astrophysics, direct capture is expected to be
important close to the driplines because of the low particle
separation energies. The cross section $\sigma_\mathrm{DC}^{\mu \nu}$
for direct capture of a projectile on a target in state $\mu$ into a
final state $\nu$ can be calculated, e.g.\ in the potential model
\cite{dr06,sat83,glen83} or in the Lane-Lynn formalism
\cite{laly}. Again, the total capture cross section is the sum of
the contributions of the different transitions to discrete states
\begin{eqnarray}
&&\sigma_\mathrm{DC} ^\mu (E_\mathrm{c.m.},E^\mu,J^\mu,\pi^\mu)=  \nonumber \\
&&\quad \sum _\nu C_\nu ^2 S_\nu \times \nonumber \\
&&\times \sigma_{\mathrm{DC}} ^{\mu \nu}
(Q+E_\mathrm{c.m.}-E^\nu,J^\mu,\pi^\mu,E^\nu,J^\nu,\pi^\nu) \, ,
\end{eqnarray}
with $C$ and $S$ denoting the isospin Clebsch-Gordan coefficient and the
spectroscopic factor of the final state, respectively. Because of the
importance of direct capture close to the driplines only, the
discrete states are usually experimentally unknown and have to be
extracted from microscopic calculations. However, it has been shown
\cite{rau98} that the resulting cross section is very sensitive to the
predicted properties of the states and that thus different microscopic
models yield vastly different results. The
sensitivity is much higher than in the case of compound reactions
because no averaged quantities are used. In order to
circumvent the problem of the exact prediction of the final states it
has been suggested \cite{raurep96,holz97,gordc97} to employ averaged
properties also for direct captures, i.e.\ to replace the sum over discrete
final states by an integration over a level density
\begin{eqnarray}
{\bar \sigma}_\mathrm{DC}^\mu&=&\int \limits _0 ^Q \sum _{J^\nu,\pi^\nu}
{\bar C}^2(E^\nu,J^\nu,\pi^\nu) {\bar S} (E^\nu,J^\nu,\pi^\nu) \times \nonumber \\
&\times& \rho(E^\nu,J^\nu,\pi^\nu) \times \nonumber \\
&\times& \sigma_\mathrm{DC}^{J^\mu \pi^\mu \rightarrow J^\nu \pi^\nu}
(Q+E_\mathrm{c.m.}-E^\nu) dE^\nu
\label{eq:dc}
\end{eqnarray}
and to use
averaged quantities $\bar C$ and $\bar S$ which are assumed as being
constant \cite{gordc97} or energy-dependent \cite{raurep96,holz97,ejn98}.
Again, the expression can be extended to include sums over thermally
populated target states $\mu$. The situation in the case of a direct
capture reaction is shown in Fig.\ \ref{fig:DC}.

The sum over spin and
parity in Eq.\ (\ref{eq:dc}) usually contains only few contributing
terms because of the dominance of E1 transitions. 
Because of the small $Q$ values the integration does not include such
a large energy range as in the Hauser-Feshbach calculations.
These two facts are also the reason
why the impact of a non-equilibrated parity ratio will be much larger
for direct cross sections than for Hauser-Feshbach ones. The
Hauser-Feshbach model assumes implicitly that the compound nucleus
can be formed with any spin and parity (although not with equal probability
for each configuration) and that the decay of the compound state by $\gamma$-
or particle-emission can proceed in transitions to a number of final states.
This offers a larger number of possible spin/parity combinations through
which the reaction can proceed than in the case of a direct process.

Summarizing, the nuclear level density is important in two aspects. Firstly,
to determine the dominant reaction mechanism. Secondly, to calculate
cross sections accounting for transitions to a number of final states.
It should be noted that 
these two aspects require the knowledge of the level density at different
excitation energies and in different nuclei, the compound nucleus in the
former case and the final nucleus in the latter. In the following section
\ref{sec:astro_results}
we study the impact of a changed parity distribution on reaction rates
calculated in the two reaction mechanisms described above.

\subsection{Comparison to standard rates and possible implications}
\label{sec:astro_results}

\subsubsection{Statistical model rates}
\label{sec:astro_hf}

Before employing the realistic parity distribution derived in 
Sec.\ \ref{sec:ratios}
in calculations of astrophysical reaction rates, we want to arrive at an
estimate of the maximum impact to be expected.
The level density enters in the calculation of the transmission
coefficient as given by Eq.\ (\ref{eq:trans}). For the standard assumption
of equally distributed parities $\rho(E^\nu,J^\nu,\pi^\nu)=\rho(E^\nu,J^\nu)/2$
for all parities $\pi^\nu$. As an extreme case we suppose that the parity
ratio is Zero at all energies, i.e.\ only states with the same parity 
$\pi^0=\pi^g$ as the
ground state occur and $\rho(E^\nu,J^\nu,\pi^\nu)=0$ for $\pi^\nu\neq\pi^g$.
This also implies that $\rho(E^\nu,J^\nu,\pi^g)=\rho(E^\nu,J^\nu)$.
We further assume that the formation
coefficient
$^kT_{y'}^{J^k\pi^k}$ dominates the total transmission coefficient in the
denominator of Eq.\ (\ref{eq:hfcs}) and thus the process $y$ and its
transmission coefficients determine the cross section. This can be further
simplified by comparing laboratory rates only, i.e.\ rates with the reaction
target being in the ground state.

Two cases can be treated separately, capture reactions and reactions with
particle emission, i.e.\ $y=\gamma$ and $y=\mathrm{n,p,}\alpha,\dots$,
respectively. In the latter case, $\pi^k$ determines the allowed partial
waves in the transition to the state in the final nucleus
with $\pi^\nu$, even partial waves
for $\pi^k=\pi^\nu$ and odd partial waves otherwise. For the capture reaction,
with $\pi^\nu$ being in the compound nucleus, the parities select the
allowed electromagnetic transition, i.e.\ E1 when $\pi^k\neq\pi^\nu$ and
M1 otherwise. For simplicity, electromagnetic
transitions with higher order are neglected. For our estimate we also assume
that only s- and p-waves contribute to the cross section significantly.

With the above assumptions, the transmission coefficient calculated with
Eq.\ (\ref{eq:trans}) will be comprised of an arithmetic mean between the
contributions of the two allowed transitions when taking equidistributed
parities. With the restriction of having only one parity available, one
of these transitions will be suppressed completely. Therefore, the
resulting transmission coefficient will be smaller or larger according to
which of the transitions became unavailable. The enhancement factor
can be expressed as
\begin{equation}
\label{eq:upfactor}
\mathcal{F}^\uparrow=\frac{2}{1+\chi} \, ,
\end{equation}
the suppression factor is found to be
\begin{equation}
\label{eq:downfactor}
\mathcal{F}^\downarrow=\frac{2\chi}{1+\chi} \, ,
\end{equation}
with $\chi\leq1$ being the ratio of p- and s-wave transmission coefficient
or of M1 and E1 transmission coefficient, respectively,
\begin{equation}
\chi = \left\{ \begin{array}{ll}
\frac{T^{\ell=1}}{T^{\ell=0}} & \textrm{for particles,}\\
\frac{T^\mathrm{M1}}{T^\mathrm{E1}} & \textrm{for radiation.}
\end{array}
\right.
\end{equation}
It can be seen immediately that the enhancement cannot exceed a factor of
two even when $\chi$ is very small. On the other hand, the suppression can
become strong depending on the ratio of the transmission coefficients.
Since usually $\chi$ is smaller for electromagnetic transitions than for
different partial waves of the particles, the largest effects are expected
for capture reactions. However, the effect depends on the
involved parities. Table \ref{tab:effect} shows the possible combinations.
\begin{table}
\caption{\label{tab:effect}Enhancement and suppression of transmission
coefficients relative to an equal parity distribution.}
\begin{ruledtabular}
\begin{tabular}{ccc}
$\pi$ & Allowed Transition & Factor\\
\hline
$\pi^k=\pi^\nu$ & $\ell=0$ & $\mathcal{F}^\uparrow$ \\
                & M1       & $\mathcal{F}^\downarrow$ \\
$\pi^k\neq \pi^\nu$ & $\ell=1$ & $\mathcal{F}^\downarrow$ \\
                & E1       & $\mathcal{F}^\uparrow$
\end{tabular}
\end{ruledtabular}
\end{table}

\begin{figure}
\includegraphics[width=6cm,angle=-90]{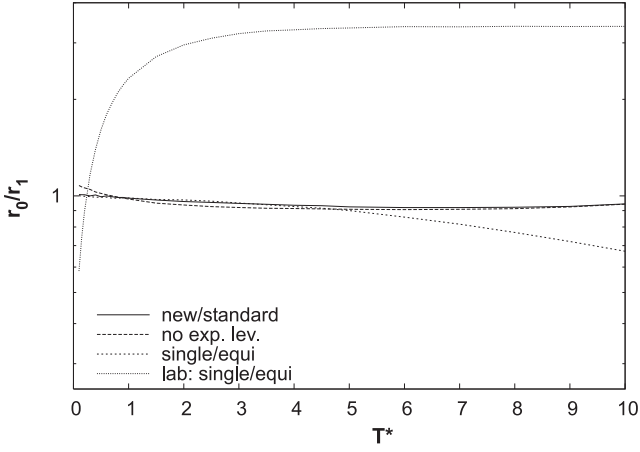}
\caption{\label{fig:se67ng}
Ratio of reaction rates as a function of stellar temperature $T^*$
(in $10^9$ K) for the reaction
$^{67}$Se(n,$\gamma$)$^{68}$Se (note the logarithmic scale);
the line denoted
"lab: single/equi" gives the ratio of laboratory rates with
the rate $r_0$
calculated assuming that only the ground state parity can appear in
a nucleus whereas an equal distribution of parities was assumed in rate
$r_1$ and neglecting any experimentally known excited states.
Similarly, the ratio of {\it stellar} rates is plotted as "single/equi".
The full line gives the ratio of the "new" stellar rate $r_0$ calculated with
the parity-dependence derived in this work and making use of the level
information as given in Table III of Ref.\ \cite{rt01} and the "standard"
stellar rate $r_1$ as published in Refs.\ \cite{rt00,rt01}. For comparison,
the ratio obtained without consideration of experimental level information is
also plotted as "no exp.\ lev." (see text).
}
\end{figure}

\begin{figure}
\includegraphics[width=6cm,angle=-90]{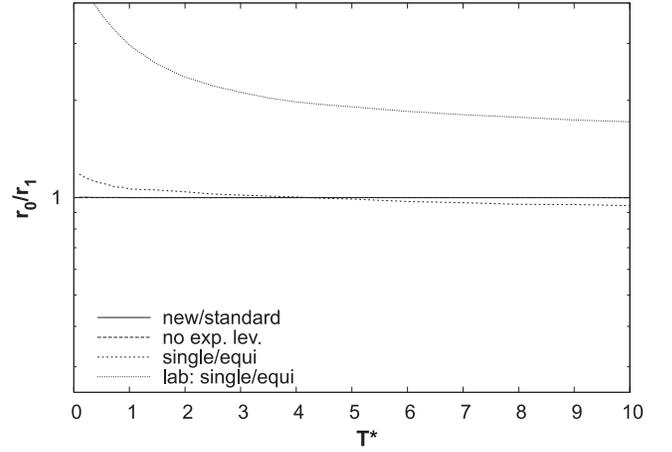}
\caption{\label{fig:se68ng}
Same as Fig.\ \ref{fig:se67ng} but for the reaction
$^{68}$Se(n,$\gamma$)$^{69}$Se
}
\end{figure}

\begin{figure}
\includegraphics[width=6cm,angle=-90]{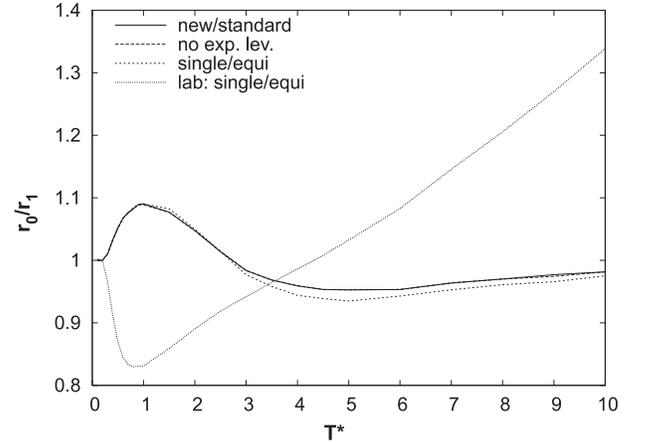}
\caption{\label{fig:se68pg}
Same as Fig.\ \ref{fig:se68ng} but for the reaction
$^{68}$Se(p,$\gamma$)$^{69}$Se; note the different scale in the figure.
}
\end{figure}

\begin{figure}
\includegraphics[width=6cm,angle=-90]{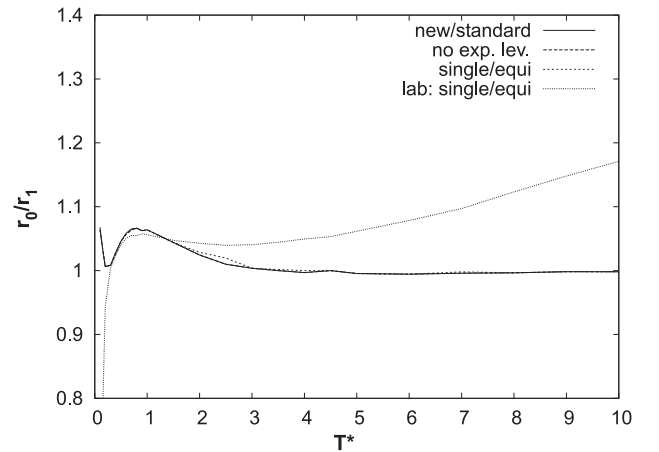}
\caption{\label{fig:se67pg}
Same as Fig.\ \ref{fig:se68pg} but for the reaction
$^{67}$Se(p,$\gamma$)$^{68}$Se
}
\end{figure}

An examination of all ratios across the nuclear chart shows that most of them
stay in the range given by the above defined $\mathcal{F}^\uparrow$ and
$\mathcal{F}^\downarrow$. However, there are several exceptions exhibiting
an enhancement of up to a factor of 5. This can be understood by
examining the partial widths of the relevant transitions which are
directly proportional to the transmission coefficients. In the derivation
above it was assumed that the total width is dominated by one of the
transitions also appearing in the numerator of Eq.\ (\ref{eq:hfcs})
and thus cancelling with the
denominator. However, if the total transmission coefficient is dominated
by other transitions, e.g.\ proton- and $\alpha$-emission in the case of a
(n,$\gamma$) reaction, then the changes in both $T_y$ and $T_{y'}$ have
to be multiplied, leading to an upper limit of a factor of 4. Considering that
also the total width may be slightly changed, the factor of 5 found in some
cases is easily accommodated. Fig.\ \ref{fig:se67ng} shows the case of
$^{67}$Se(n,$\gamma$) where the neutron width of the compound state $k$ is
larger than the $\gamma$ width but much smaller than both the proton- and
$\alpha$ widths. The situation is slightly different for $^{68}$Se(n,$\gamma$),
shown in Fig.\ \ref{fig:se68ng}, where the $\gamma$ width is much larger
than the neutron width but considerably smaller than the proton- and $\alpha$
widths.
If the ratio would be fully dominated by one process,
then $\mathcal{F}^\uparrow$ and 
$\mathcal{F}^\downarrow$ should increase and decrease, respectively,
with increasing plasma temperature because
at higher excitation energy of the compound nucleus more intermediate levels
can be reached. Thus, either more levels can be populated by the preferred
transition, giving rise to $\mathcal{F}^\uparrow$, or more transitions are
missing for $\mathcal{F}^\downarrow$.
In fact, the total effect stems from an interplay between the
stellar temperature dependences of the formation and decay widths (and thus
of the enhancement and suppression factors) and their ratios.
This can lead to a complicated temperature dependence of the rate ratios
with initially declining and finally increasing ratios or vice versa.
Examples are shown in Figs.\ \ref{fig:se68pg} and \ref{fig:se67pg}.

The above examples are extreme cases,
including a comparison between rates with a parity ratio of zero and rates
with equidistributed parities. To show more pronounced effects, no
experimentally known levels in the participating nuclei were considered and the
targets were assumed to be in the ground state. For more realistic
astrophysical reaction rates, the targets have to be thermally excited
according to the plasma temperature which increases the number of
possible channels for the formation of the compound nucleus. Moreover,
all statistical model calculations consider experimentally determined
low-lying states, employing a theoretical level density only above the
last included level. This will further reduce the effect of a modified
parity distribution but only for nuclei close to stability for which
level information is available. Moreover, higher partial waves and
multipolarities than considered
in the previous estimate will also contribute, thereby further washing out the
effects seen above. Figs.\ \ref{fig:se67ng}--\ref{fig:se67pg} also show the
actual ratio obtained when comparing a full calculation of stellar rates
employing the parity distribution derived in Sec.\ \ref{sec:ratios} and including
experimental information from Table III of \cite{rt01} with the standard FRDM
rate set published in \cite{rt00,rt01}.

\subsubsection{Direct capture}
\label{sec:astro_dc}

In principle, Eqs.\ (\ref{eq:upfactor}) and (\ref{eq:downfactor}) also apply for
the enhancement and suppression factors, respectively, in the direct capture
mechanism. As already mentioned in Sec.\ \ref{sec:astro_models}, however, the
transitions allowed by the spin and parity selection rules
are more limited than in the compound model
and thus the application is more straightforward. Furthermore, there is no
energy-dependent renormalization by a total width. Finally, due to the
nature of the electromagnetic multipole operators appearing in the expression
for the direct capture cross section, E1 radiation is dominating by far
and all other multipoles are strongly suppressed (see, e.g., \cite{rau98}).
These factors will lead
to a higher sensitivity of direct captures to the parity ratio.

We defer a large-scale calculation of direct capture rates for low
capture $Q$ values to another paper but rather want to illustrate the
above by a simplified example here. Assuming neutron capture on a $\pi_A=+1$
target state the allowed neutron partial wave $\ell _\alpha$ in the
entrance channel is determined by the parity selection rule
\begin{equation}
(-1)^{\ell _\alpha} \pi_A =(-1)^{\ell _\gamma} \pi_B \quad .
\end{equation}
Since E1 ($\ell _\gamma=1$) emission is by far dominating, this can be
further simplified to
\begin{equation}
(-1)^{\ell _\alpha}=-\pi_B \quad ,
\end{equation}
with $\pi_B$ being the parity of the final state. Therefore only even partial
waves will be allowed for $\pi_B=-1$ and only odd partial waves for $\pi_B=+1$.
The latter relation would just reverse if the target state had odd parity.
Because of the low projectile energies relevant in nuclear astrophysics,
only s- and p-waves will have a significant contribution to the cross section.
Thus, the ratio $\chi$ to be inserted in Eqs.\ (\ref{eq:upfactor}) and (\ref{eq:downfactor}) becomes
\begin{equation}
\chi=\frac{\sigma^{\mu \rightarrow \nu,\textrm{p-wave}}_\mathrm{DC}}{\sigma^{\mu \rightarrow \nu,\textrm{s-wave}}_\mathrm{DC}} \quad ,
\end{equation}
with $\chi<1$ because of the higher s-wave cross section ($\sigma_\mathrm{s} \propto 1/\sqrt{E}$) at small energy.
Because of the limited availability of alternative transitions, contrary to
the Hauser-Feshbach case the suppression or enhancement cannot be compensated
in a direct reaction and the factors $\mathcal{F}^\uparrow$ and
$\mathcal{F}^\downarrow$ will have their full effect on each transition.
For example, $\chi=0.18$ for capture of a 100 keV neutron
on the $0^+$ ground state of
$^{140}$Sn and ending in a final $3/2$ state with an energy release of
1 MeV (see also \cite{rau98}).
The inclusion of thermally populated target states will slightly reduce
the effect by making available states with different spin but same parity
because of a predominant parity at low excitation energy. This will
have a smaller impact than in the Hauser-Feshbach model since low
partial waves dominate the cross section, there is no moderating effect of
a total width, and generally fewer possible transitions.
\begin{table}
\caption{\label{tab:ffdc}Enhancement and suppression of transmission
coefficients relative to an equal parity distribution for direct capture
on a target state with parity $\pi^\mu_A$ and ending in a final state
with parity $\pi^\nu_B$. The factors $\mathcal{F}^\uparrow$ and
$\mathcal{F}^\downarrow$ are defined in Eqs.\ (\ref{eq:upfactor}) and
(\ref{eq:downfactor}), respectively.}
\begin{ruledtabular}
\begin{tabular}{ccc}
$\pi$ & Allowed Transition & Factor\\
\hline
$\pi^\mu_A=\pi^\nu_B$ & $\ell _\alpha=1,3,5,\dots$ & $\mathcal{F}^\downarrow$ \\
$\pi^\mu_A\neq \pi^\nu_B$ & $\ell _\alpha=0,2,4,\dots$ & $\mathcal{F}^\uparrow$
\end{tabular}
\end{ruledtabular}
\end{table}

The application rule for $\mathcal{F}^\uparrow$ and
$\mathcal{F}^\downarrow$ for direct capture is shown in Tab.\ \ref{tab:ffdc}.
For the case of the final nucleus in a capture reaction having opposite
parity than the target ground state,
the capture rate will be enhanced relative to a rate calculated
using a level density with equally distributed parities
because of the predominance of states with ground state parity at low
excitation energies.
This may only
be altered if an overshooting of the opposite parity occurs, as discussed
in Sec.\ \ref{sec:ratio_results}. However, direct capture is important
only at such low separation energies that the overshooting regime
is not reached.

\subsubsection{Discussion}
\label{sec:astro_disc}

We have calculated parity distributions for all nuclei between proton and neutron
dripline from Ne to Bi and recomputed the Hauser-Feshbach
astrophysical reaction rates given
in \cite{rt00,rt01} with the same parameters as quoted there (set FRDM).
Selected examples were already shown in Figs.\
\ref{fig:se67ng}--\ref{fig:se67pg}. In Sec.\ \ref{sec:astro_hf} we discussed
considerable enhancement or suppression for certain transitions. In the
astrophysical context, however, the impact of the new parity ratio remains
limited. This is due to several reasons.

Firstly, the impact of a change in the parity distribution is higher for
reactions with large $Q$ values because a larger number of transitions from
the compound nucleus is energetically accessible.
For example, neutron captures will be
altered more strongly on the proton-rich side whereas capture on neutron-rich
targets -- provided the compound reaction mechanism is dominating -- will be
less affected. A similar observation can be made for proton capture. Thus,
the proton captures occuring in the $rp$ process \cite{schatzpr,schatzend}
and the neutron captures in the $r$ process \cite{ctt91,qw96,frr99,mey89}
show changes of several percent relative to the rates computed with an
equal distribution of parities.
\begin{figure}
\includegraphics[width=6cm,angle=-90]{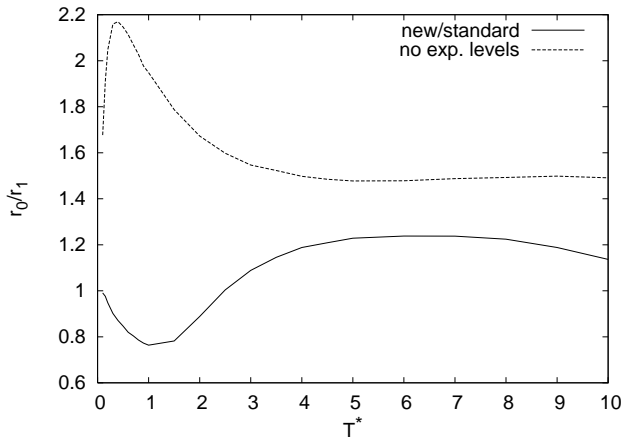}
\caption{\label{fig:nb94pg}
Stellar reaction rate ratios for $^{94}$Nb(p,$\gamma$)$^{95}$Mo as function
of stellar temperature $T^*$ (in units of 10$^9$ K); plotted
is the ratio of the stellar rate $r_0$ calculated with the parity
distribution derived in this work with the standard rate $r_1$ from
\cite{rt00,rt01}. Further plotted is the ratio of the same rates but
without inclusion of experimentally known low-lying levels (see text).
}
\end{figure}
\begin{figure}
\includegraphics[width=6cm,angle=-90]{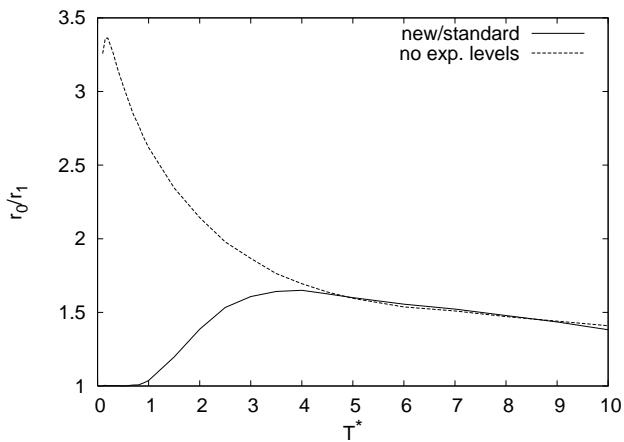}
\caption{\label{fig:zr95pg}
Same as Fig.\ \ref{fig:nb94pg} but for the reaction
$^{95}$Zr(p,$\gamma$)$^{96}$Nb
}
\end{figure}

Moreover, for small capture $Q$ values, the statistical model will not be
applicable and resonant or direct capture will dominate. We expect
a larger impact of the parity distribution in the averaged direct capture
model but further conclusions have to await the recalculation of the
direct rates. Even with changed rates, however, it has to be kept in mind
that the processes occurring close to the driplines are often in equilibrium,
where individual rates are not important to determine the produced nuclear
abundances \cite{r05}. Rates will only become important in the freeze-out
phases when the nucleosynthesis path moves closer to stability
\cite{schatzpr,frr99,r05}. In these phases the compound mechanism will
set in again.

Close to the valley of stability
level schemes are known to comparatively high excitation energies and
are used in the calculations of the relevant nuclear cross
sections. Usually, up to 20 levels are included whenever known \cite{rt00,rt01}.
This will further reduce the effect of the applied parity distribution
because an effective, non-equilibrated parity distribution is already in
use for the lowest excitation energies. Figs.\ 
\ref{fig:se67ng}--\ref{fig:se67pg} show a comparison of rates with and
without included experimental level information. It should be noted that
the slight change in $^{67}$Se(n,$\gamma$) is not due to neglected levels
in $^{68}$Se but rather due to the modified transmission coefficients in the
competing channels, i.e.\ proton- and $\alpha$ emission. In the example
shown in Figs.\ \ref{fig:nb94pg} and \ref{fig:zr95pg},
20 low-lying levels have been considered
in all channels in the standard calculation. Removing them, the impact of
the parity distribution when using pure level densities can be seen. Generally,
only few low-lying states have to be included because the transitions with
the highest relative energy, leading to the lowest states,
dominate in all channels. Due to the inclusion of experimental information
in the standard calculations, the impact of a changed parity distribution
is also small for the late freeze-out phases and the nucleosynthesis phases
involving nuclei close to the valley of stability. This includes the
photodisintegration reactions in $p$-process nucleosynthesis
\cite{wh78,ray90,arngor}, even though
they occur on a longer timescale than any freeze-out process.

Finally, the most important mechanism suppressing the impact of the parity
distribution is the necessity to include transitions on excited states in
the target. In an astrophysical plasma, excited states can be thermally
populated as specified in Eq.\ (\ref{eq:csstar}). 
Because of the Bohr independence hypothesis for compound reactions
this allows for a large
number of alternative transitions in case one of the parities is suppressed.
Small alterations of the energy-dependence of the cross sections
cannot be seen in the astrophysical reaction rates anymore due to the
folding with the Maxwell-Boltzmann energy distribution of the interacting
nuclei according to Eq.\ (\ref{eq:rr}). The fact that the impact of the
parity dependence is strongly suppressed in {\it stellar} rates can be
clearly seen in Figs.\ \ref{fig:se67ng}--\ref{fig:zr95pg}. The larger
the stellar temperature, the smaller the parity impact.

Concluding, inclusion of the parity distribution in this work leads to a
change in {\it stellar} reaction rates of only several percent up to about 10\%
in astrophysically relevant reactions. A few cases exhibit changes up to a
factor of 2 or 1/2, respectively, but are not astrophysically important.

\section{Summary}
\label{sec:sum}

We generalized a method to calculate the excitation-energy dependent parity
distribution in nuclei and used it to calculate parity ratios for all
nuclei up to Bi. We demonstrated the importance of including a
sufficiently large single-particle model space. In particular, excitations
from core single-particle states can also be important.
The parity ratio proved to be non-equilibrated up to 5--15 MeV. Interestingly,
an overshooting effect in the parity ratio was found at major shell closures
where states with ground-state parity are outnumbered by states of
opposite parity over a range of several MeV.

The derived parity distribution was then used to recalculate
astrophysical reaction rates. The impact of the new description on
stellar rates including thermal excitations of the target proved to be
limited in the Hauser-Feshbach model
although the impact on direct capture has to be studied further.

\acknowledgments
This work was supported in part by the Swiss Nationalfonds (grants 2124-055832,
2000-061031, 2000-105328) and by the U.S. DOE (grant DE-FG-0291-ER-40608).

\end{document}